\documentclass[%
 amsmath,amssymb,
pra,
twocolumn
]{revtex4-2}

\usepackage{graphicx}
\usepackage{dcolumn}
\usepackage{bm}

\usepackage{subcaption}
\usepackage{tikz}
\usepackage{xcolor}
\usepackage{hyperref}
\hypersetup{
 pdfauthor={},
 pdftitle={},
 pdflang={English},colorlinks=true,linkcolor=cyan,citecolor=teal,urlcolor=cyan}

\begin{document}

\preprint{APS/123-QED}

\title{Collision Dynamics of Bose-Einstein Condensates}

\author{Aaron Wirthwein}
\author{Stephan Haas}%
    \affiliation{Department of Physics and Astronomy, University of Southern California, Los Angeles, CA 90007}
    
\author{Sheng-wey Chiow}
    \affiliation{Jet Propulsion Laboratory, California Institute of Technology, Pasadena, CA 91109, USA}

\date{\today}

\begin{abstract}
We study the collision dynamics of two Bose-Einstein condensates with their dynamical wave functions modeled by a set of coupled, time-dependent Gross-Pitaevskii equations. Beginning with an effective one-dimensional system, we identify regimes characterized by the relationship between inter- and intra-atomic interactions and the initial configuration of the system, akin to the equilibrium phase diagram of two interacting Bose condensates. We consider a dynamical setup in which two wave packets are initially at rest with a small separation about the center of an asymmetric harmonic trap. Upon release, we observe a rapid approach to dynamical equilibrium in the limits of very large and very small inter-particle repulsion, characterized by periodic transmission or reflection of the condensates as distinguishable units, whereas the intermediate, critical regime is characterized by extended transient dynamics, density fracturing, and dynamical mixing. We therefore have reason to believe that non-trivial behavior may exist during the collisions of Bose-Einstein condensates as a result of their non-linear interactions, and these effects may be observable in a suitable laboratory environment.
\end{abstract}

\maketitle

\section{Introduction}
An environment which is free from limitations imposed by gravity could enable the study of phenomena that would otherwise be difficult or impossible to realize in a traditional lab setting. In the field of cold atomic gases, experiments involving multi-component Bose-Einstein condensates on the ground often suffer from differential gravitational sagging and limited times of flight \cite{Frye2021}. Conducting such experiments in space has the potential to eliminate the effects of gravity and enable extended flight times, and indeed this has been demonstrated aboard an Earth-orbiting facility using condensates composed of rubidium atoms \cite{Aveline2020}.  The ability to suspend condensates without regard for their masses or internal magnetic states would provide an ideal setting for the study of long term collision dynamics between condensates composed of different atoms. 

An interesting feature of binary BECs is that they exhibit a range of ground state configurations depending on the atomic interactions. Binary BECs exhibit miscibiliy or immiscibility depending on the relative strength of these interactions. As observed in experiments \cite{Myatt1997, Hall1998, Lee2018, McCarron2011, Pasquiou2013, Tojo2010, Wang2015}, when the interspecies repulsive interaction is large enough, the two components repel each other so that they separate into two distinct clouds with small spatial overlap, corresponding to the immiscible phase. When the effect of interspecies interactions is weak compared to the intraspecies interactions, the two components are miscible and overlap with each other at the center of the trapping potential.

It is well known for a homogeneous binary BEC, where the kinetic energy is negligible compared with the interaction energy, the transition between the miscible and immiscible phases can be determined by the interspecies and intraspecies interactions. The traditional criterion for the immiscible ground state is $ a_{12}^2 > a_{11}a_{22} $ \cite{Pitaevski2003, Smith2001}, where $ a_{11} $ and $ a_{22} $ are the intraspecies s-wave scattering lengths of components 1 and 2, while $ a_{12} $ is the interspecies s-wave scattering length. One can then control the miscibility of the gases by using the Feshbach resonance technique to adjust the interaction strengths \cite{Tojo2010, Wang2015}. However, if some asymmetry is introduced in the trapping potential, atomic masses, or atom number, the traditional boundary set by $ a_{12}^2 = a_{11}a_{22} $ no longer holds \cite{Wen2020}. 

Research on binary BECs has been primarily concerned with static ground state properties, most likely due to the difficulty in achieving binary mixtures experimentally for sufficiently long duration. There have been a few studies which investigated dynamical behavior \cite{Hall1998, Maddaloni2000}, but they are only capable of studying dynamics over very short time scales ($\approx10$ ms) and for small kinetic energies. Our studies suggest that interesting behavior may exist in the collisions of binary BECs over long time evolutions ($ > 1s $) and for relatively large kinetic energies. We study a system composed of two fully-condensed clouds of ultracold Rubidium atoms in different hyperfine states confined to an asymmetric harmonic trap. 


We begin by introducing a theoretical model for condensate evolution in an asymmetric harmonic trap and outline the static ground state phases for comparison to the dynamical scenario we study in this paper. Then we describe our computational modeling of condensate evolution using a split-step Fourier method (SSM) over a finite spatio-temporal grid. Our results are then presented in a way that illustrates the dyanimal phases we observe by varying the trap frequency and inter-atomic interaction strength. Our methods are then outlined in more detail, and we finish with a discussion including the future outlook for the study of condensate collisions in microgravity.

\section{Theoretical model}
We consider two condensates composed of 1000 $^{87}\mathrm{Rb}$ atoms in the hyperfine spin states $|1\rangle=|1,-1\rangle$ and $|2\rangle=|2,+1\rangle$. Let $a_{11}$ be the s-wave scattering length of a BEC composed of atoms in the $|1\rangle$ state, and $a_{22}$ be the s-wave scattering length for $|2\rangle$. There is also an interatomic interaction between the condensates parameterized by $a_{12}$. Experimentally derived values of the scattering lengths are given in Table \ref{table:simulation-parameters} from \cite{Egorov2013}. The condensates reside in an asymmetric harmonic trap:
\begin{equation}
V(\mathbf{r}) = \frac{1}{2}m\left(\omega_{x}^{2}x^{2}+\omega_y^2 y^{2}+\omega_{z}^{2} z^{2}\right),
\label{eqn:3d-trap-potential}
\end{equation}
where $m$ is the mass of rubidium. The trap is designed with the $x-y$ confinement being strong enough to create a quasi one-dimensional condensate in the $z$ direction, but not strong enough to violate the Born approximation for the pseudo-potential of the atomic interactions. That is, if the radial trap frequency is given by $\omega_r = \omega_x = \omega_y,$ then the trap energy $\epsilon_r = \hbar \omega_r$ is large enough that the condensate remains in the ground state of the trap in the $x-y$ plane, but is free to explore new states in the $z$ direction. Additionally, we assume that if $\ell_r = \sqrt{\hbar/ m \omega_r}$ is the characteristic harmonic oscillator length, and $a_s$ is the s-wave scattering length of the condensate, then the Born approximation requires $a_s/\ell_r\ll 1,$ which is monitored and strictly enforced in our simulations.

The effective one-dimensional dynamics of two rubidium BECs occupying different hyperfine states can be described within a mean-field model by the following set of time-dependent, coupled Gross-Pitaevskii equations \cite{Pitaevski2003}:
\footnotesize
\begin{subequations}
\begin{align}
    i \hbar \frac{\partial \psi_{1}}{\partial t} &= \left(-\frac{\hbar^{2} \nabla^{2}}{2m}+V+N g_{11}\left|\psi_{1}\right|^{2}+N g_{12}\left|\psi_{2}\right|^{2}\right) \psi_{1},\\
    i \hbar \frac{\partial \psi_{2}}{\partial t} &= \left(-\frac{\hbar^{2} \nabla^{2}}{2 m}+V+N g_{22}\left|\psi_{2}\right|^{2}+N g_{12}\left|\psi_{1}\right|^{2}\right) \psi_{2},
\end{align}
\label{eqn:time-dependent-gpe}
\end{subequations}
\normalsize
where $N$ is the atom number, and $\psi_1$ and $\psi_2$ are the 1D wavefunctions describing the condensates of atoms in $|1\rangle$ and $|2\rangle$ respectively. Here we've also introduced re-normalized interaction parameters as a result of the radial confinement:
\begin{equation}
    g_{ij} =\frac{2\hbar^{2} a_{i j}}{m \ell_{r}^{2}}
\end{equation}
where $i,j$ refer to intra-atomic $i=j$ and inter-atomic $i\neq j$ interaction scattering lengths.

\begin{table*}[t]
\begin{tabular}{|cc|cc|cc|cc|}
\hline
\multicolumn{2}{|c|}{Spatio-temporal grid}      & \multicolumn{2}{c|}{Scattering lengths} & \multicolumn{2}{c|}{Trapping frequencies} & \multicolumn{2}{c|}{Cloud composition} \\ \hline
Box size       & 320 $\mu$m                     & $a_{11}$          & $100.40a_0$         & $\omega_x$      & $2\pi\times 100$        & Atomic mass         & 85.4678 u        \\
Grid points    & $10^4$                         & $a_{22}$          & $95.44a_0$          & $\omega_y$      & $2\pi\times 100$        & Atom number         & $10^3$           \\
Evolution time & $40\times\frac{2\pi}{\omega_z}$ & $a_{12}$          & $98.006a_0$         & $\omega_z$      & $2\pi\times (5-50)$     &                     &                  \\
Trotter steps  & $10^6$                         &                   &                     &                 &                         &                     &                  \\ \hline
\end{tabular}
\caption{Table of parameters used in numerical simulations of colliding condensates composed of rubidium atoms.}
\label{table:simulation-parameters}
\end{table*}

When released at a fixed separation relative to the trap minimum, we expect the condensates to oscillate and collide multiple times over the course of their evolution. We seek to understand how the inter-atomic interaction energy $g_{12}$ and the trap frequency $\omega_z$ affect the ensuing dynamics. Before moving to the dynamical simulations, let's first discuss the static ground state phases, which will serve as a point of reference for the dynamical phases discussed later on. Beginning from the energy functional \cite{Pitaevski2003},
\begin{align}
\begin{split}
    E=\int d z &\bigg\{\sum_{j=1,2}\left[\frac{\hbar^{2}}{2 m}\left|\partial_{z} \psi_{j}\right|^{2}+\frac{1}{2} m \omega_{z}^{2} z^{2}\left|\psi_{j}\right|^{2}\right]+\\
   &\frac{g_{11}}{2}\left|\psi_{1}\right|^{4}+\frac{g_{22}}{2}\left|\psi_{2}\right|^{4}+g_{12}\left|\psi_{1}\right|^{2}\left|\psi_{2}\right|^{2}\bigg\},
\label{eqn:energy-functional}
\end{split}
\end{align}
we apply a variational method for finding approximate ground state properties. In this work, we consider the two condensates to be initially un-mixed and take a form closely resembling the ground state of the trap, which is approximately Gaussian in nature. We therefore take the following Gaussian ansatz to perform our analysis: 
\begin{equation}
    \psi_{j}=\pi^{-1 / 4} w_{j}^{-1 / 2} e^{-\left(z-z_{j}\right)^{2} /\left(2 w_{j}^{2}\right)},
\label{eqn:ansatz}
\end{equation}
where $j = 1,2$, $w_j$ is a variational parameter related to the width of the Gaussian, and $z_j$ is another variational parameter representing the central location of the Gaussian distribution. Inserting this ansatz into Eqn.~(\ref{eqn:energy-functional}), we have
\begin{align} 
\begin{split}
E =& \frac{1}{4}N m \omega_z^2(w_1^2+2z_1^2) + \frac{1}{4}N m \omega_z^2(w_2^2+2z_2^2) \\
&+\frac{1}{4}\sqrt{\frac{2}{\pi}} \frac{N^2 g_{11}}{w_1} + \frac{1}{4}\sqrt{\frac{2}{\pi}}\frac{N^2 g_{22}}{w_2} \\  
&+ \frac{\hbar^2}{4m} \left( \frac{N}{w_1^2} + \frac{N}{w_2^2} \right) + \frac{N N g_{12}}{\sqrt{\pi}} \frac{e^{-\delta^2/(w_1^2+w_2^2)}}{\sqrt{w_1^2+w_2^2}},
\end{split}
\label{eqn:energy-ansatz}
\end{align}
where we've defined $\delta = z_1 - z_2.$ For the variational step, we now minimize Eqn.~(\ref{eqn:energy-ansatz}) with respect to $z_i, w_i$ to obtain the following conditions for the ground state configuration:
\begin{subequations}
\begin{gather}
\begin{split}
    \bigg(1-\frac{1}{w_{i}^{4}}-&\frac{N g_{i i}}{\sqrt{2 \pi}} \frac{1}{w_{i}^{3}}\bigg)=\\
    &\frac{2 N g_{12}}{\sqrt{\pi}} \frac{w_{1}^{2}+w_{2}^{2}-2 \delta^{2}}{\left(w_{1}^{2}+w_{2}^{2}\right)^{5 / 2}} e^{-\frac{\delta^{2}}{w_{1}^{2}+w_{2}^{2}}},
\end{split}\\
    \delta\left[1-\frac{4 N g_{12}}{\sqrt{\pi}} \frac{e^{-\delta^{2} /\left(w_{1}^{2}+w_{2}^{2}\right)}}{\left(w_{1}^{2}+w_{2}^{2}\right)^{3 / 2}}\right]=0
\end{gather}
\label{eqn:ground-state-params}
\end{subequations}
We can make a few simplifying assumptions to arrive at a more intuitive understanding by setting $g_{11} = g_{22} = g.$ In this limit, $w_1 = w_2 = w$ and if we further restrict our attention to small $\delta$,
\begin{equation}
    w^{3}=\sqrt{\frac{2}{\pi}} \frac{N\left(g+g_{12}\right)}{2 m \omega_{z}^{2}},
\label{eqn:gs-width}
\end{equation}
and
\begin{equation}
    \delta = \sqrt{2}\left(\frac{2 N}{\pi} \frac{g+g_{12}}{2}\right)^{1 / 3} \sqrt{\log \left[\frac{2 g_{12}}{g+g_{12}}\right]}.
\label{eqn:delta-static}
\end{equation}
We see evidence of a transition at $g_{12} = g.$ For $g_{12} > g,$ the solution is given approximately by Eqn.~(\ref{eqn:delta-static}), and for $g_{12} < g, $ $\delta = 0.$ This transition is continuous in the sense that $\delta$ approaches the same value from either side of the transition. More sophisticated numerical techniques are required to arrive at a full solution to Eqns.~(\ref{eqn:ground-state-params}), but we have gleaned a few key insights from the analysis presented thus far. From Eqn.~(\ref{eqn:gs-width}), it is clear that the widths of the density clouds should increase with increasing atom number and atomic interactions, and decrease with increasing trap frequency. It is interesting to note that Eqn.~(\ref{eqn:delta-static}) does not appear to depend on the linear trap frequency. This may be the result of our simplifying assumptions, or it could be the result of the symmetry inherent to our harmonic trapping potential. It is well-known that asymmetry, whether it be in atomic masses, atom number, or applied potential, can alter the traditional criteria for the immiscibility transition. 

\section{Computational modeling}

\begin{figure}
    \centering
    \includegraphics[width=0.95\columnwidth]{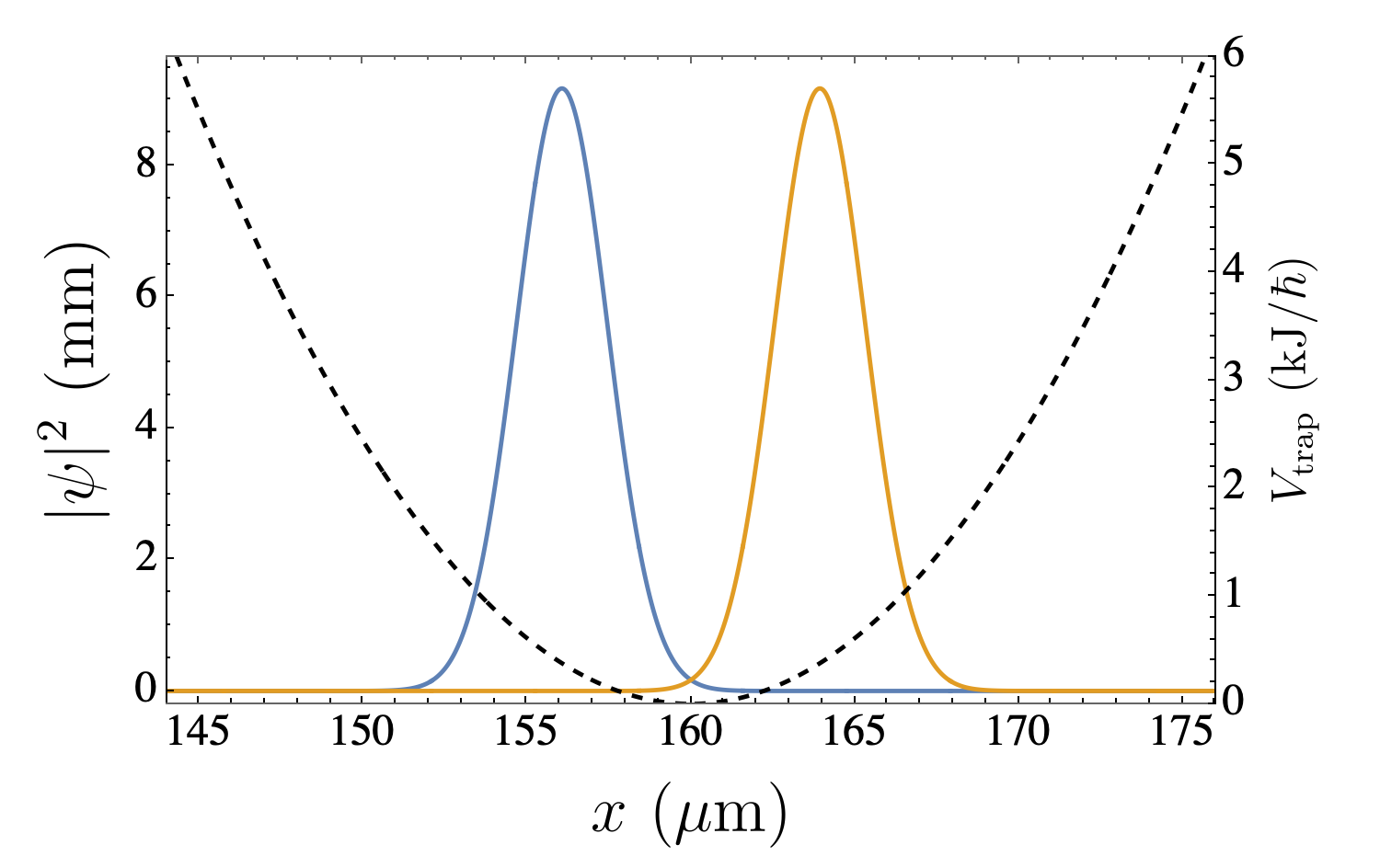}
    \caption{Initial configuration for two BECs in a parabolic trap. The density profile pictured on the right models $^{87}$Rb in the $|1,-1\rangle$ hyperfine state with native s-wave scattering length of $a_{11} = 100.40a_0,$ where $a_0$ is the Bohr radius. On the left is the density profile of $^{87}$Rb in the $|2,1\rangle$ state with s-wave scattering length $a_{22} = 95.44a_0.$ The condensates are separated by $7.8~{\mu {\rm m}}$ symmetrically about the center of a harmonic trap with frequency $\omega_z = 2\pi*30~{\rm Hz}.$}
\label{fig:initial-config}
\end{figure}

Now we turn our attention to dynamical evolution of the two condensates in the trap. Using a split-step Fourier method (SSM) (see section \ref{methods}), we solve Eqns.~(\ref{eqn:time-dependent-gpe}) over a finite spatio-temporal grid (with parameters outlined in Table \ref{table:simulation-parameters}). The system is initialized with each condensate taking a Gaussian form having width equal to the oscillator length $\ell_z = \sqrt{\hbar/m\omega_z},$ and central locations displaced symmetrically about the trap center by a distance of $4\ell_z.$ This separation was chosen to ensure minimal initial overlap without introducing excessive kinetic energies throughout the evolution. We choose to scale the separation with the trap frequency in this way to have approximately equivalent kinetic energies upon impact at the trap center. A plot of the initial configuration for a trap frequency of $2\pi\times 30 $ Hz is shown in Fig.~\ref{fig:initial-config}. After being released, the condensates are drawn towards the center of the trap and deform slightly in response to intra-atomic interactions. In the limit of negligible kinetic energy, the condensates would take on a shape given by the Thomas-Fermi approximation \cite{Smith2001}. As they approach the center of the trap, the inter-atomic interactions increase with increasing overlap. We expect the condensates to partially transmit and/or reflect in a way that depends on the magnitude of the inter-particle interactions. 

The SSM can be used to solve for the evolution of the condensates in the appropriate limit for which the Gross-Pitaevski equation can be applied. From the initial configuration, we use the SSM to propagate the wavefunction according to the Eqns.~(\ref{eqn:time-dependent-gpe}). As the condensates evolve, we track their central location as well as the second and third moments relating to the width and skewed-ness of their density distributions respectively. We vary the interaction constant over the expected critical regime and the trapping frequency from 5-50 Hz.

\section{Results}

\begin{figure*}
     \centering
     \begin{subfigure}{.4\textwidth}
         \centering
         \begin{tikzpicture}
             \draw (0, 0) node[inner sep=0] {\includegraphics[width=\textwidth]
             {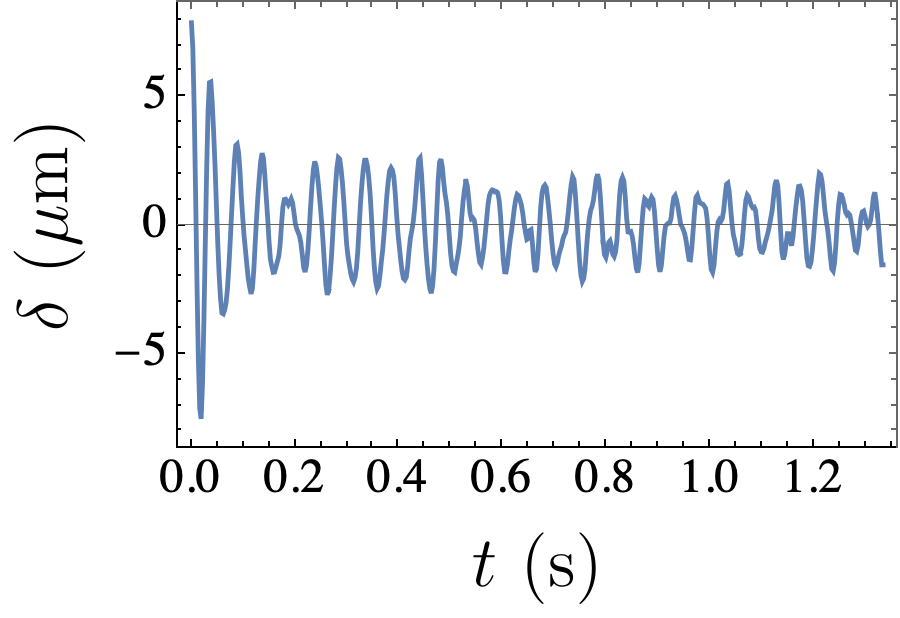}};
             \draw (3, 2) node {(a)};
        \end{tikzpicture}
         \label{fig:r1}
     \end{subfigure}
     \hfill
     \begin{subfigure}{.4\textwidth}
         \centering
         \begin{tikzpicture}
             \draw (0, 0) node[inner sep=0] {\includegraphics[width=\textwidth]
             {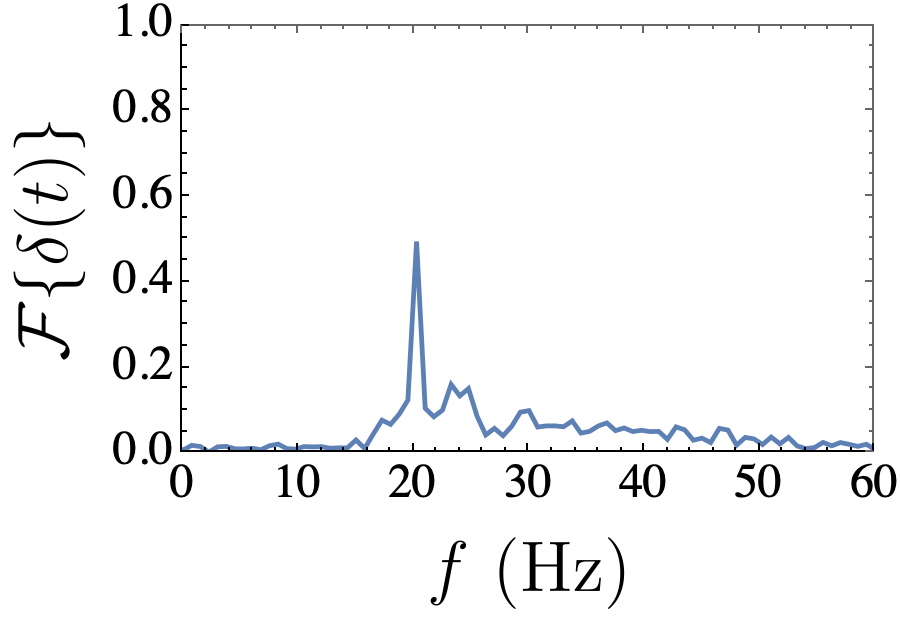}};
             \draw (3, 2) node {(b)};
         \end{tikzpicture}
         \label{fig:r1f}
     \end{subfigure}
     \newline
     \begin{subfigure}{.4\textwidth}
         \centering
         \begin{tikzpicture}
             \draw (0, 0) node[inner sep=0] {\includegraphics[width=\textwidth]
             {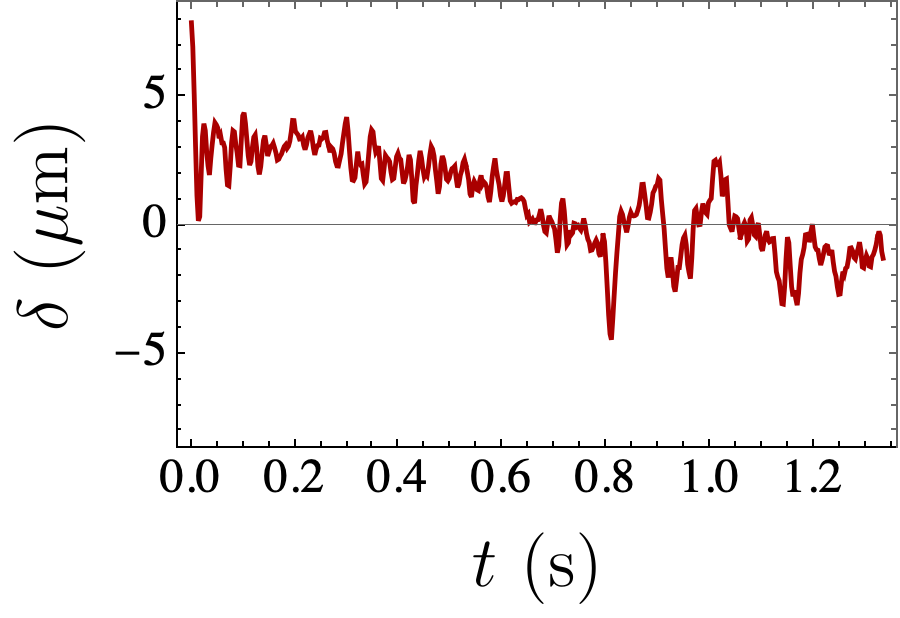}};
             \draw (3, 2) node {(c)};
         \end{tikzpicture}
         \label{fig:r2}
     \end{subfigure}
     \hfill
     \begin{subfigure}{.4\textwidth}
         \centering
         \begin{tikzpicture}
             \draw (0, 0) node[inner sep=0] {\includegraphics[width=\textwidth]
             {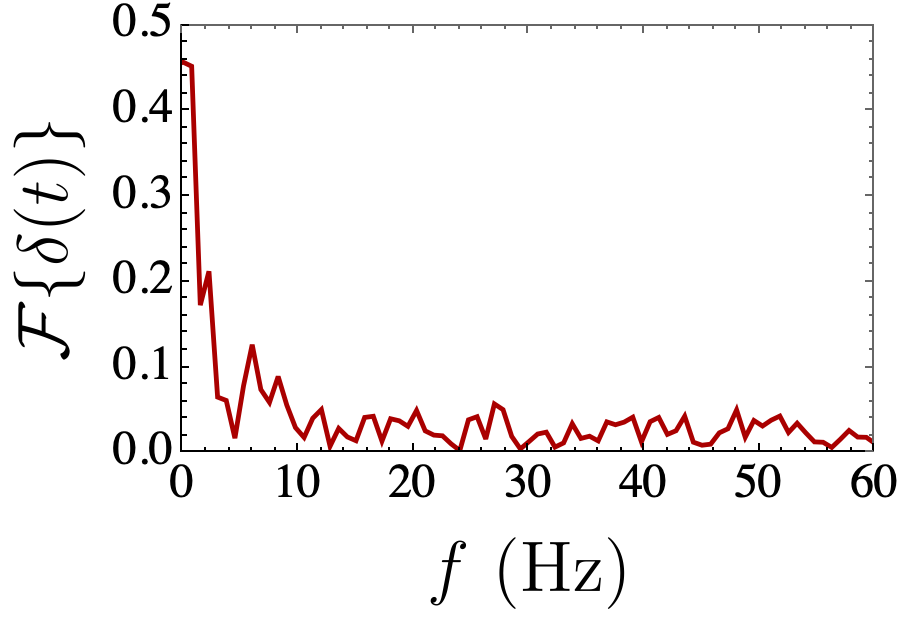}};
             \draw (3, 2) node {(d)};
         \end{tikzpicture}
         \label{fig:r2f}
     \end{subfigure}
     \newline
          \begin{subfigure}{.4\textwidth}
         \centering
         \begin{tikzpicture}
             \draw (0, 0) node[inner sep=0] {\includegraphics[width=\textwidth]
             {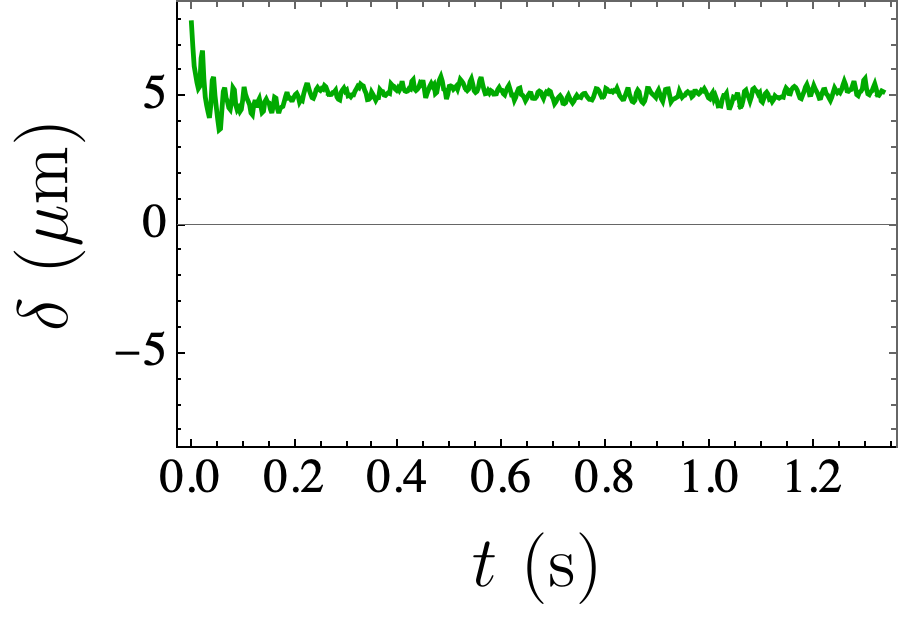}};
             \draw (3, 2.1) node {(e)};
         \end{tikzpicture}
         \label{fig:r3}
     \end{subfigure}
     \hfill
     \begin{subfigure}{.4\textwidth}
         \centering
         \begin{tikzpicture}
             \draw (0, 0) node[inner sep=0] {\includegraphics[width=\textwidth]
             { 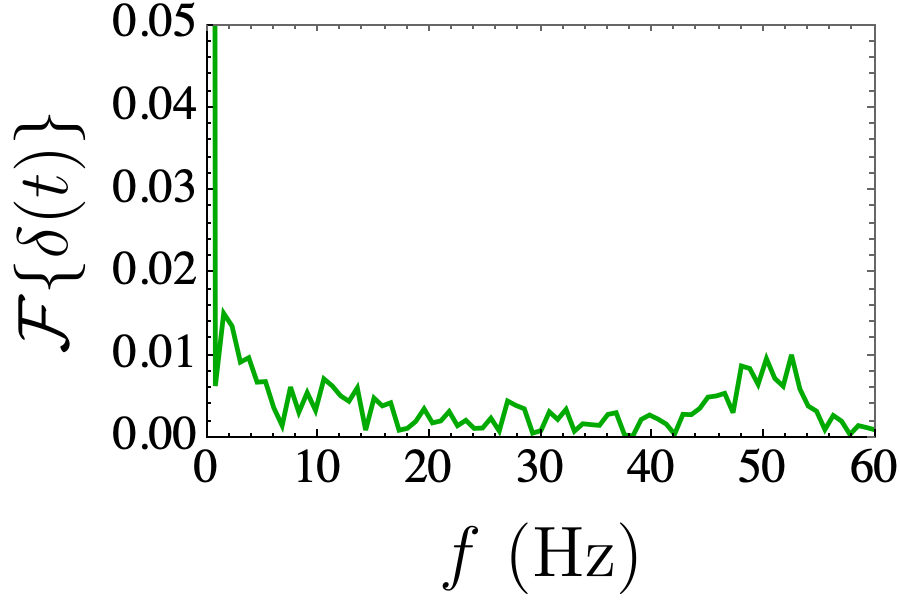}};
             \draw (3, 2) node {(f)};
         \end{tikzpicture}
         \label{fig:r3f}
     \end{subfigure}
\caption{(a,c,e) Raw data from SSM output of $\delta$ vs time for three different interaction energies. (b,d,f) Fourier transform of $\delta(t)$ in the corresponding plot to the left. These data come from three simulations using a trap frequency of 30 Hz, and interaction energies of (a-b) $g_{12}$ = 3.26 $\mu$m/s, (c-d) $g_{12}$ = 8.47 $\mu$m/s, and (e-f) $g_{12}$ = 16 $\mu$m/s.}
\label{fig:raw-data}
\end{figure*}
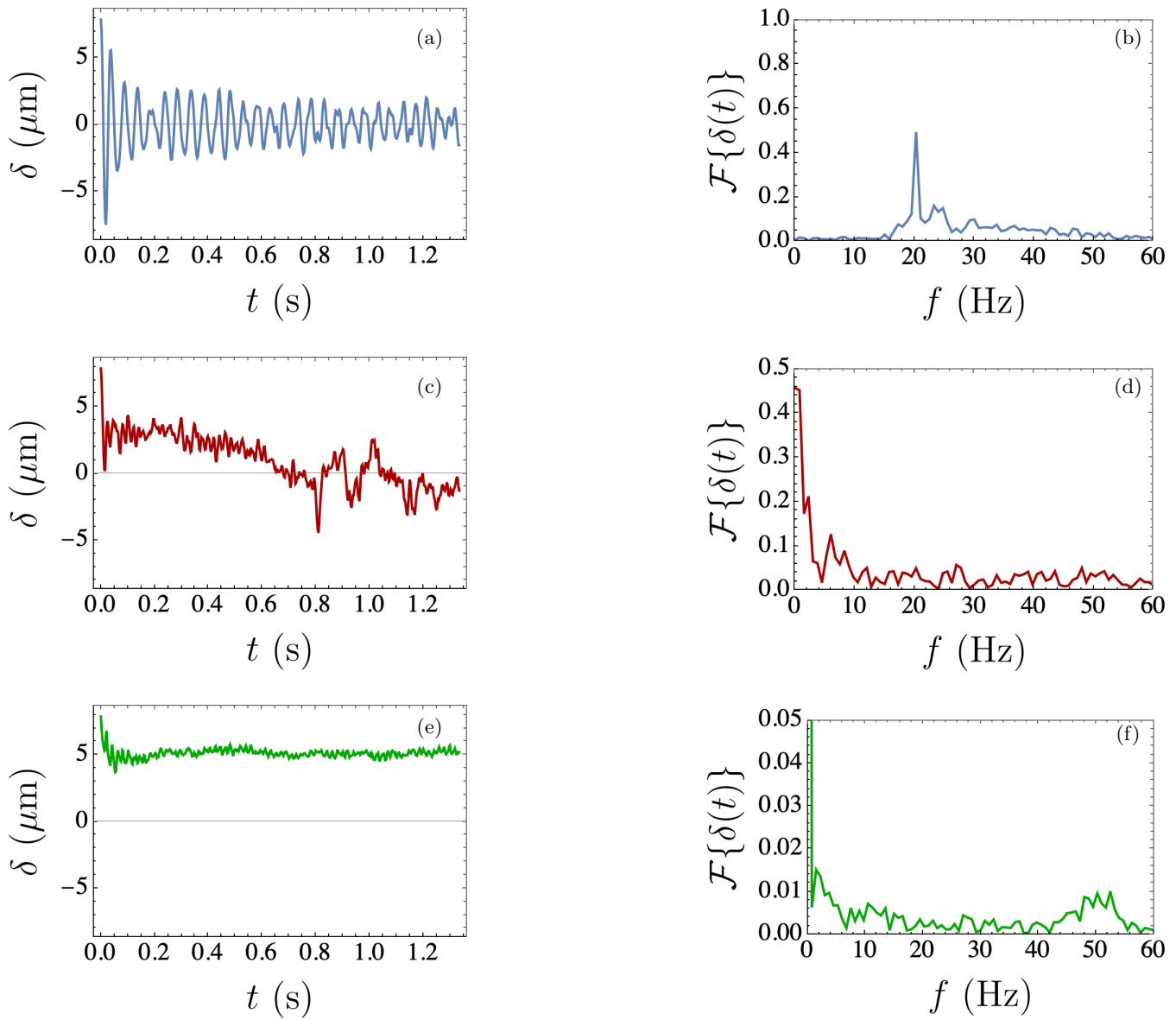

While the time evolution is calculated using $10^6$ time steps, the output of our SSM is $\psi_{1,2}$ over 500 representative samples in time. We use this output to calculate the mean position and higher order moments of $\psi$. Let $\delta$ denote the difference in position between the right and left condensate, then 
\begin{equation}
    \delta = \int~dz (|\psi_1|^2-|\psi_2|^2) z, 
\end{equation}
and furthermore define
\begin{equation}
    \mu_m(\psi_i) = \int~dz |\psi_i|^2 z^m, 
\end{equation}
as the higher order moments.  Fig.~\ref{fig:raw-data} illustrates some of the raw data for $\delta$ we collected in three simulations using a trap with frequency of 30 Hz. In the case of relatively small or large interactions, we observe a rapid approach to a dynamical equilibrium characterized by a well-defined frequency of motion. 

In the regime with small interactions, the condensates mostly transmit upon collision with minor slowing in response to an energy exchange between kinetic and interaction terms. This is evidenced in Fig.~\ref{fig:raw-data}(a-b). The primary mode of oscillation can be extracted from the Fourier transform of $\delta(t)$, and we find for nonzero interactions, the oscillation frequency is generally smaller than the trap frequency in this regime. 

As we move closer to the critical regime, we observe diverging transient times. Fig.~\ref{fig:raw-data}(c-d) illustrates a typical example of a system with interaction strength near the expected critical value. There is no clear sign of approach to dynamical equilibrium over the specified time evolution. It is possible that at a later time, this system may settle into a mode of oscillation, but it is unclear from the present work if this is the case. We observe diverging transient times in this regime over a broad range of trap frequencies, justifying the claim that the system does indeed exhibit critical behavior.

For even larger interactions, we observe yet another rapid approach to a different kind of oscillation characterized by periodic reflections of the condensates, or ``bouncing." The dominant feature in the corresponding Fourier spectrum (Fig.~\ref{fig:raw-data}f) indicates the existence of an average, and there appears to be a sub-leading term which captures the frequency of the bouncing mode. This frequency is generally larger than the trap frequency, although never quite double, as would be expected in the case of infinite repulsion at the trap center.

\begin{figure}[t]
     \centering
     \begin{subfigure}[b]{0.8\columnwidth}
         \begin{tikzpicture}
             \draw (0, 0) node[inner sep=0] {         \includegraphics[height=.8\textwidth]
             { 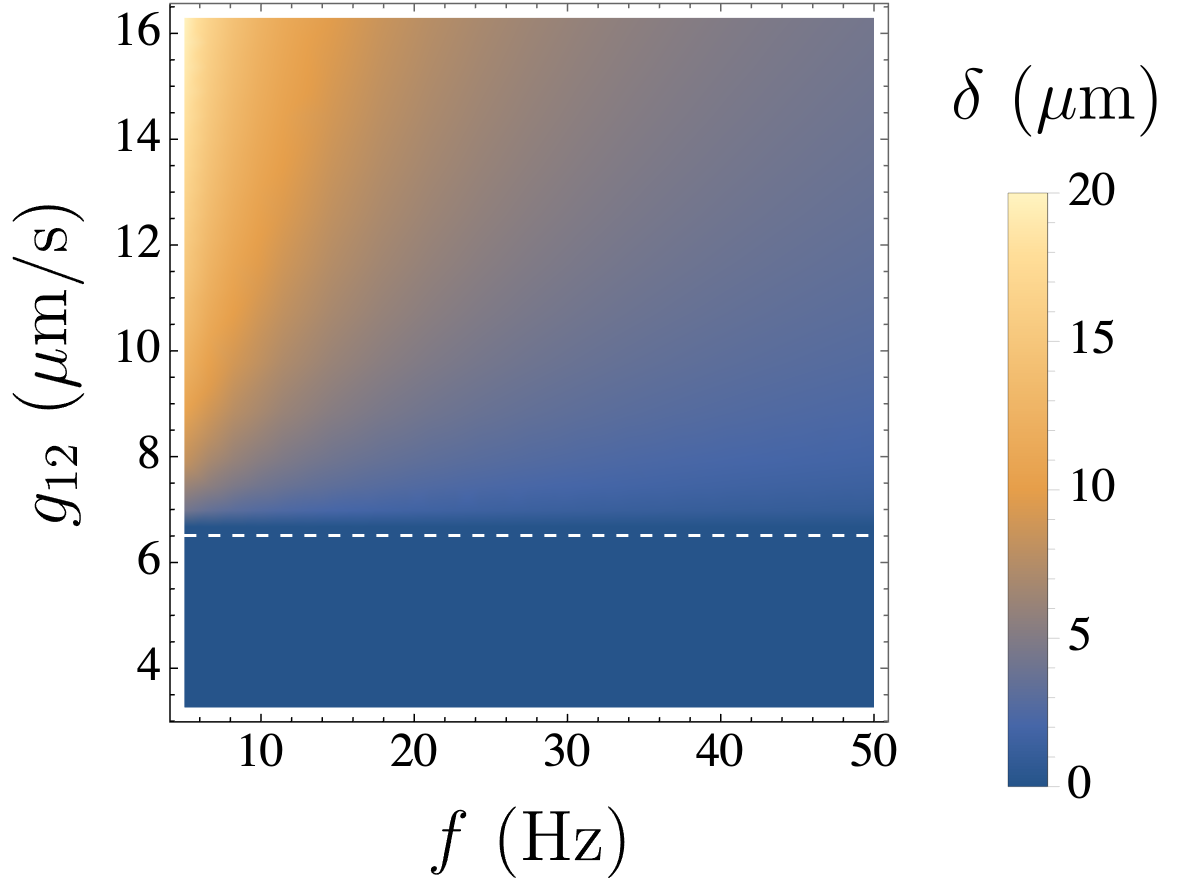}};
             \draw (1.4, -1.25) node[text=white] {(a)};
         \end{tikzpicture}
         \label{fig:delta-static}
     \end{subfigure}
     \hfill
     \begin{subfigure}[b]{0.8\columnwidth}
         \centering
         \begin{tikzpicture}
             \draw (0, 0) node[inner sep=0] {         \includegraphics[height=.8\textwidth]
             { 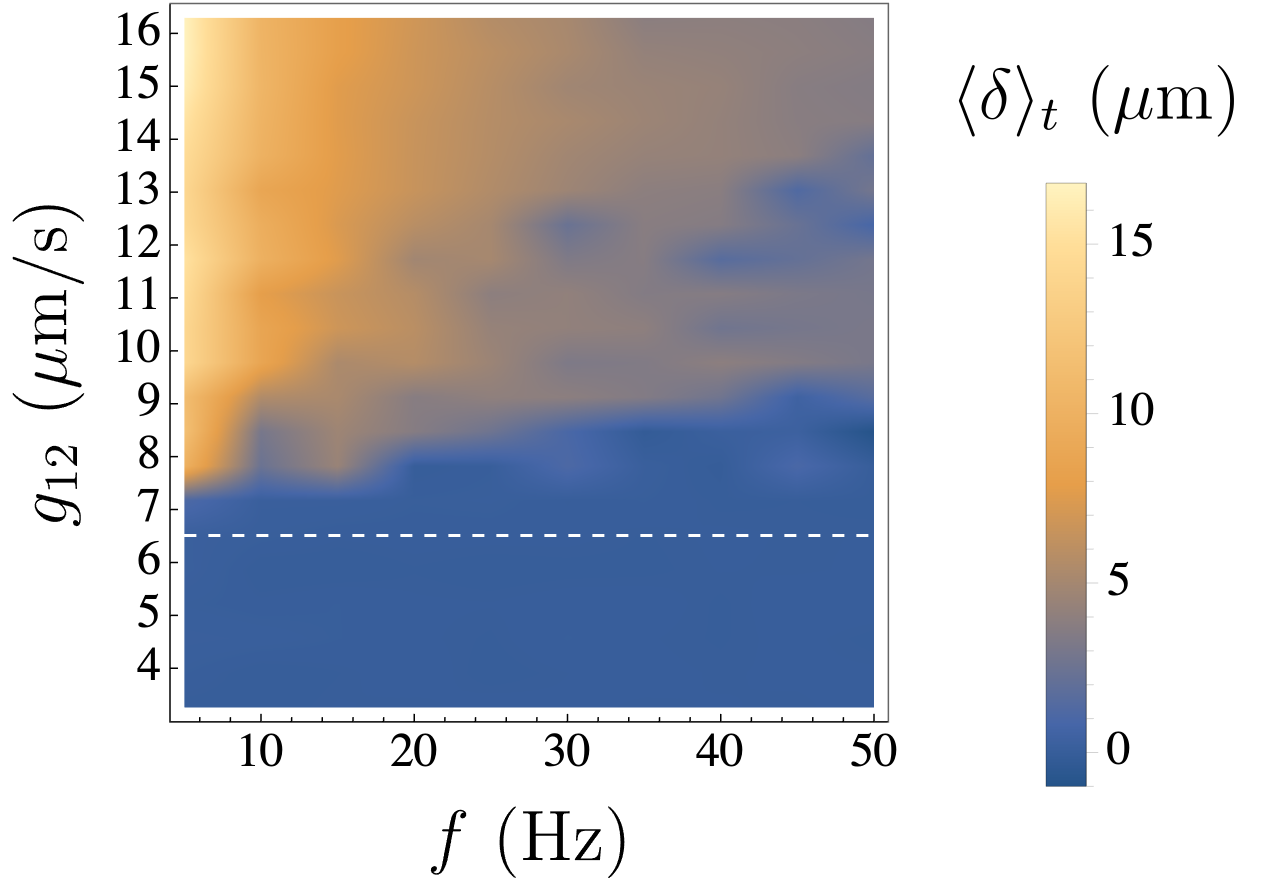}};
             \draw (1.15, -1.25) node[text=white] {(b)};
         \end{tikzpicture}
         \label{fig:delta-dynamic}
     \end{subfigure}
\caption{Density plots showing center of mass separation as a function of inter-particle interaction $g_{12}$ and trapping frequency $f$ for static (a) and dynamical (b) settings. The dotted white line indicates the critical value of $g_{12} = \sqrt{g_{11} g_{22}}.$ }
\label{fig:delta-phases}
\end{figure}

We perform similar analyses for a range of interactions and trapping frequencies. By taking the time average of $\delta(t)$ for each simulation, we can compare our dynamical results to the corresponding static case, ie the ground state of two interacting Bose-Einstein condensates. The results are summarized in Fig.~\ref{fig:delta-phases}. There are a few key similarities between the two plots. We see in both systems there is zero average separation for interactions below the critical value, and the magnitude of the separations follow similar trends generally speaking; As the interaction increases, so do the separations in both cases. The separation also decreases in response to stronger trapping frequencies as well. The first key difference to point out is that there seems to be an alteration in the critical line for the dynamic setting. There seems to be some dependence on the trapping frequency which is not evident in the static case. For stronger traps, it would appear that the critical point exists at some value $g_{12} > \sqrt{g_{11}g_{22}}.$

\begin{figure}[t]
     \centering
     \begin{subfigure}[b]{.47\textwidth}
         \centering
         \begin{tikzpicture}
             \draw (0, 0) node[inner sep=0] {\includegraphics[width=\textwidth]
             { 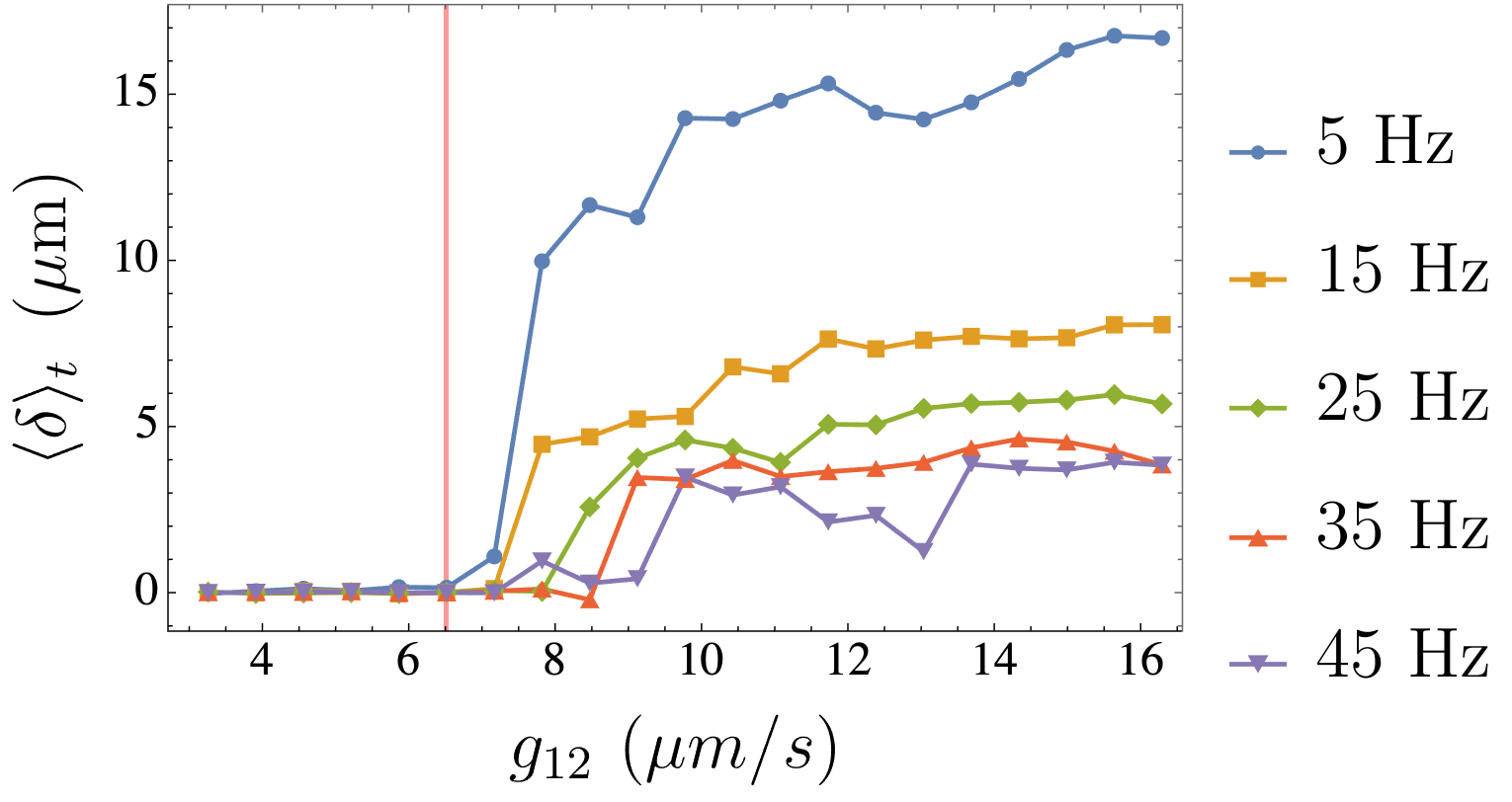}};
             \draw (-2.5, 1.5) node {(a)};
         \end{tikzpicture}
         \label{fig:m1}
     \end{subfigure}
     \hfill
     \begin{subfigure}[b]{.47\textwidth}
         \centering
         \begin{tikzpicture}
             \draw (0, 0) node[inner sep=0] {\includegraphics[width=\textwidth]
             { 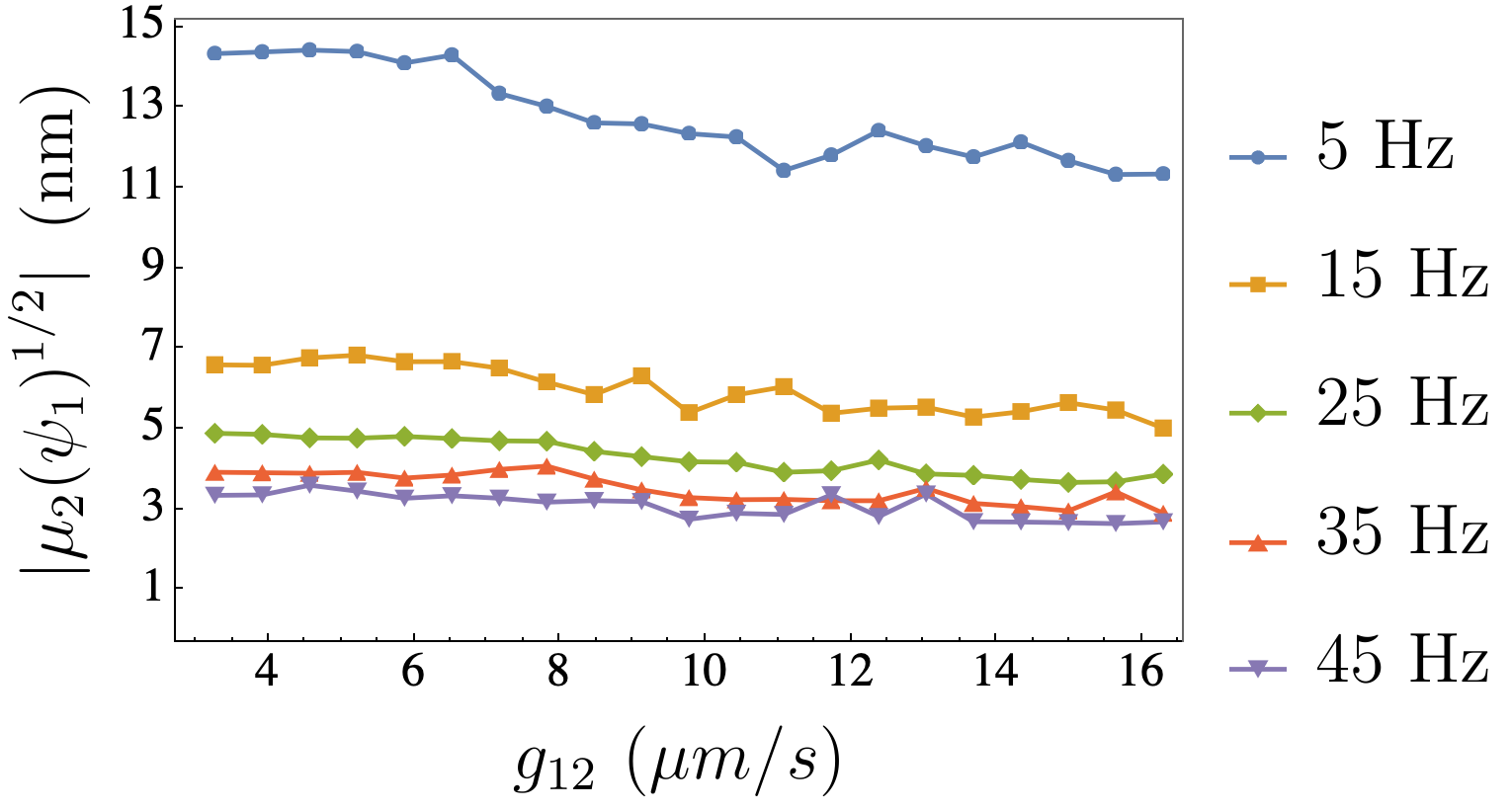}};
             \draw (-2.5, 1.5) node {(b)};
         \end{tikzpicture}
         \label{fig:m2}
     \end{subfigure}
     \newline
     \begin{subfigure}[b]{.47\textwidth}
         \centering
         \begin{tikzpicture}
             \draw (0, 0) node[inner sep=0] {\includegraphics[width=\textwidth]
             { 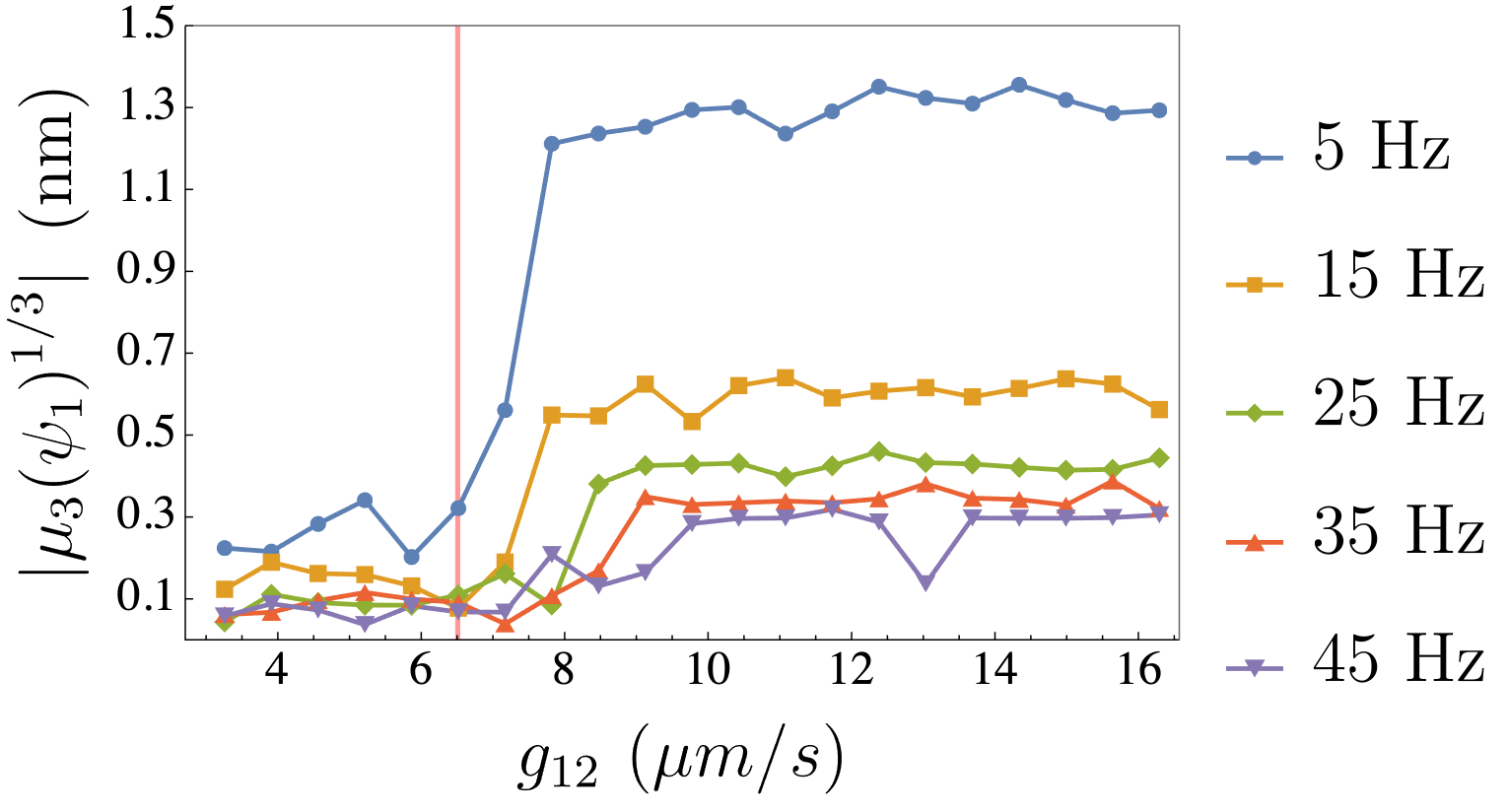}};
             \draw (-2.5, 1.5) node {(c)};
         \end{tikzpicture}
         \label{fig:m3}
     \end{subfigure}
\caption{Moments of the condensate density profile as a function of inter-atomic interaction for various trapping frequencies. (a) Time-averaged difference between first moments, or mean positions, of the condensate densities. (b) Time-averaged second moment of the $^{87}$Rb($|2,1\rangle$) condensate density profile $|\psi_1(t)|^2$. (c) Time-averaged third moment of $|\psi_1(t)|^2$. The red line in (a)~$\&$~(c) indicates the traditional critical value of $g_{12}$.}
\label{fig:moments}
\end{figure}
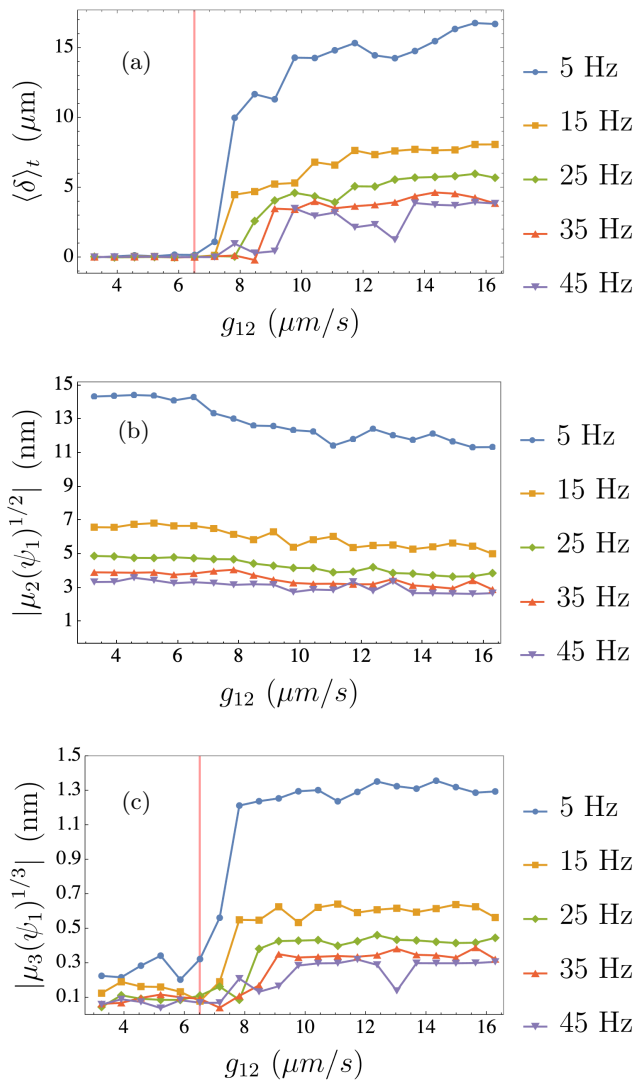

Another method of analysis involving the higher order moments leads us to further insights regarding the configuration of the binary mixture as a result of dynamical evolution in the trap. Fig.~\ref{fig:moments} contains plots of the first three moments (time-averaged) versus interaction strength for a selection of trap frequencies. While Fig.~\ref{fig:moments}(a) contains mostly the same information discussed in Fig.~\ref{fig:delta-phases}, perhaps we can more clearly see the dependence of the transition on trapping frequency. Again, for larger trapping frequencies it would appear that the transition shifts to larger values. The third moment pictured in Fig.~\ref{fig:moments}(c) also shows signs of the transition. This can be understood by noting that the third moment addresses the asymmetry of the distribution, and for large interactions, the condensates are repelling each other with a force great enough to induce deviations from the symmetric Gaussian structure it began with. The second moment on the other hand shows no sign of a transition in $g_{12}.$ This can be understood as the second moment captures the width of the distribution, which is more strongly dependent on the intra-atomic interactions and the trapping frequency. As we increase the trapping frequency, $\ell_z$ becomes smaller, and so the condensates are initialized with smaller widths. It is also interesting to note the slight decrease in width corresponding to an increase in interaction strength. This stands in opposition to the analysis presented for the static case, and indeed the resolution to this discrepancy has not yet been uncovered. 

\section{Methods}
\label{methods}

The work presented here relies on a pseudo-spectral numerical method used to solve nonlinear partial differential equations like Eqns.~\ref{eqn:time-dependent-gpe}. The method relies on computing the solution in small steps, and treating the linear and the nonlinear steps separately. It is necessary to Fourier transform at each step because the linear step is made in the momentum basis and the nonlinear step is made in position basis. This method is superior to other finite-difference schemes in that it is unconditionally stable, time reversible, conserves particle number, and is time-translationally invariant \cite{Bao2013}. 

To explain how the SSM works, let's first discuss the time dependent wavefunction for a single particle moving in a time-independent, one-dimensional potential. The formal solution to the Schr\"{o}dinger equation is 
\[ |\psi(t)\rangle=\exp \left[-\frac{\mathrm{i}\left(t-t_{0}\right)}{\hbar} \hat{\mathcal{H}}\right]\left|\psi\left(t_{0}\right)\right\rangle. \]
The simplest way of computing this propagation is to express the wavefunction and the Hamiltonian in a
particular basis and use matrix exponentiation to find the time dependence of the expansion coefficients. Calculating $|\psi(t) \rangle$ this way is not very efficient because matrix exponentiation is a numerically difficult operation for general Hamiltonians. A much more efficient method can be achieved through the use of incredibly fast discrete Fourier transform algorithms and an approximation involving the Trotter expansion:
\[ e^{\lambda(T+V)}\approx e^{\frac{\lambda}{2}V} e^{\lambda T}e^{\frac{\lambda}{2}V}e^{\mathcal{O}(\lambda^3)}. \]
If we set $\lambda = \frac{i (t-t_0)}{M\hbar}$ for some large $M$, we find
\begin{align}
\begin{split}
    |\psi(t)\rangle&=e^{M \lambda \hat{\mathcal{H}}}\left|\psi\left(t_{0}\right)\right\rangle=\left[e^{\lambda \hat{\mathcal{H}}}\right]^{M}\left|\psi\left(t_{0}\right)\right\rangle,\\&=\left[e^{\lambda(\hat{T}+\hat{V})}\right]^{M}\left|\psi\left(t_{0}\right)\right\rangle, \\&= \lim _{M \rightarrow \infty}\left[e^{\frac{\lambda}{2} \hat{V}} e^{\lambda \hat{T}} e^{\frac{\lambda}{2} \hat{V}}\right]^{M}\left|\psi\left(t_{0}\right)\right\rangle.
\end{split}
\end{align} 
We then represent the potential in the position basis and the kinetic operator in the momentum basis so that both matrices are diagonal. The great advantage of this approach is that algebra with diagonal matrices is as simple as algebra with scalars but applied to the diagonal elements one-by-one. So if we represent the wavefunction in the appropriate basis when acted on by either the potential or kinetic operator, then the exponentiation becomes trivial. The non-trivial step is to Fourier transform the wavefunction between it's basis representations, but this can be dealt with using fast methods that exist to calculate discrete Fourier transforms. The wavefunction at a time $t+\delta t$ can be found from the wavefunction at time $t$ via 

\begin{gather*}
        \psi(x,t +\delta t) = \\ e^{-i\frac{\delta t}{2}\hat{V}(x)}\mathcal{F}^{-1}\left\{e^{-i\delta t \hat{T}(k)}\mathcal{F}\left[e^{-i\frac{\delta t}{2}\hat{V}(x)}\psi(x,t)\right]\right\}.
\end{gather*}

The split step method can even be extended to time-dependent Hamiltonians so long as the potential varies slowly enough to be considered constant over the course of a Trotter step $\Delta t/M$. For most scenarios this approximation becomes exact as $M\to \infty.$ 

\section{Discussion}
Collision dynamics of condensate mixtures has not
been the focus of experimental research to date, and we
believe this may be partially due to the limitations imposed by gravity. From our numerical simulations, we
have reason to believe that non-trivial behavior may exist during such collisions as a result of the nonlinear interactions, and that these effects may be observable in
a suitable laboratory environment. Our results call for
long evolution times, the ability to suspend disparate atomic clouds in proximity to one another, and the ability to tune their interactions via Feshbach resonances. We propose that the microgravitational environment afforded by BECCAL may be the ideal setting for an experimental realization. 

Our numerical simulations of colliding rubidium BECs in an asymmetric harmonic trap indicate dynamical signatures of a possibly modified miscibility phase transition. The transition seems to occur at values $g_{12}>\sqrt{g_{11}g_{22}}$ which increase with greater trap frequency. Three dynamical phases have been characterized by the relative strength of atomic interactions. More work is needed to identify the boundaries and characteristics of these phases. In future work, we plan to continue working towards that effort, as well as expanding the current capabilities of our SSM to tackle systems in 2- and 3-dimensions for the study of new phases resulting from rotational degrees of freedom, and more exotic condensate geometries.

\section{Acknowledgements}
 The research was carried out in part at the Jet Propulsion Laboratory, California Institute of Technology, under a contract with the National Aeronautics and Space Administration (80NM0018D0004). The authors would like to thank the Caltech Student-Faculty Program and Dr. Lorenzo Campos Venuti for their financial support during the summer of 2021. Additionally, we acknowledge Dr. Aiichiro Nakano and Dr. Matteo Sbroscia for their consultation in this research.

\nocite{*}

\bibliography{bib}

\end{document}